# Controlling the Atomic Reconstruction in Twisted $MoSe_2/WSe_2$ with a Monolayer hBN Spacer

*Johannes Figueiredo, Kaiyuan Chen, Aubrey Penn, Mirco Troue, Sebastian Loy, Kenji Watanabe, Takashi Taniguchi, Frances M. Ross[‡], Julian Klein[†], Alexander W. Holleitner**

*J. Figueiredo, M. Troue, S. Loy, A. W. Holleitner*
Walter Schottky Institute and Physics Department, TU Munich, Am Coulombwall 4a, 85748 Garching, Germany
* Email Address: holleitner@wsi.tum.de

*J. Figueiredo, M. Troue, S. Loy, A. W. Holleitner*
Munich Center for Quantum Science and Technology (MCQST), Schellingstr. 4, 80799 Munich, Germany

*J. Figueiredo, K. Chen, F. M. Ross, J. Klein*
Department of Materials Science and Engineering, Massachusetts Institute of Technology, Cambridge, Massachusetts 02139, United States
[†, ‡] Email Addresses: jpklein@mit.edu, fmross@mit.edu

*A. Penn*
MIT.nano, Massachusetts Institute of Technology, Cambridge, Massachusetts 02139, United States

*K. Watanabe*
Research Center for Electronic and Optical Materials, National Institute for Materials Science, Tsukuba 305-0044, Japan

*T. Taniguchi*
Research Center for Materials Nanoarchitectonics, National Institute for Materials Science, Tsukuba 305-0044, Japan

## Abstract

Twisted van der Waals heterobilayers are a promising material platform for studying collective exciton many-body phenomena, but near-aligned structures often exhibit a local, anisotropic atomic reconstruction that compromises lateral homogeneity. The present study describes how atomic-scale rearrangement in adjacent regions of hBN-encapsulated, near-aligned $MoSe_2/WSe_2$ heterobilayers is controlled by the presence of a mono- or bilayer hBN spacer between the two monolayers. High-angle annular dark-field STEM directly resolves the atomic alignment in both layers and shows that the spacer suppresses the reconstruction and strongly increases the structural uniformity of the heterobilayers. 4D-STEM quantitatively maps local rotation, dilation, and heterostrain, revealing that the separated regions remain consistent with rigid moiré superlattice simulations, whereas without the spacer the structural parameters have two to three times broader distributions. By suppressing reconstruction, ultrathin hBN spacers therefore provide an effective route to enhance lattice homogeneity in $MoSe_2/WSe_2$ heterobilayers targeting collective excitonic and optoelectronic phenomena.

## Introduction

Van der Waals heterostructures composed of semiconducting transition-metal dichalcogenide (TMD) monolayers provide a versatile platform for controlling band offsets, interlayer hybridization, and excitonic properties through stacking, relative alignment, dielectric environment, and carrier density [1,2]. In TMD heterobilayers, any residual twist and/or lattice mismatch gives rise to a large-scale moiré superlattice that is associated with moiré excitons, resonantly hybridized excitonic resonances, and electrically controlled exciton dynamics [2–7]. In heterobilayers with type-II band alignment, electrons and holes can reside in different monolayers and form spatially indirect excitons [8]. For such interlayer excitons, the suppressed radiative recombination leads to lifetimes on the order of tens to hundreds of nanoseconds [9]. This long lifetime makes these excitons particularly suitable for studying transport and trapping processes, as well as gate-tunable excitonic many-body phenomena [10,11]. Optical access to these excitons is especially strong in nearly crystallographically aligned bilayers, where the electron and hole valleys remain close to momentum matching. In this regime, the moiré superlattice, which modulates the potential landscape for interlayer excitons, is highly sensitive to any structural changes [12].

A rigid-lattice moiré description, however, is not generally sufficient in the low-twist-angle limit. As in other stacked van-der-Waals systems (e.g. bilayer graphene), in TMD bilayers with comparable lattice constants, the competition between stacking-dependent interlayer energy and intralayer elastic energy can drive atomic reconstruction, due to the strong dependence of the interlayer interaction on local atomic registry between upper and lower layers [13,14]. As a result, rather than sampling all local registries between layers smoothly across the moiré cell, the bilayer relaxes to form enlarged low-energy domains separated by narrow domain walls [15–17].

This atomic reconstruction has become established as a characteristic structural response of near-aligned TMD bilayers with small lattice mismatch. Any reconstruction, by definition, involves local strain, rotations and dilations, whose effects on the bilayer properties must be considered beyond relative translation alone [18]. This departure from a rigid moiré geometry is directly relevant for optical and electronic properties of the bilayers, because the local stacking configuration governs interlayer hybridization, modifies the single-particle band landscape, and shapes the effective potential experienced by interlayer excitons [3–6,19,20]. Optical signatures associated with moiré trapping or miniband formation must therefore be interpreted in the context of the relaxed lattice rather than the nominal twist angle alone [21–23]. In particular, a structural, quantitative description is essential for interpreting excitonic phenomena in reconstructed moiré TMD bilayers, where localization and related many-body effects are often discussed in terms of an assumed simple moiré potential [1,24].

A possible route to controlling the excitonic landscape is to separate the two TMD monolayers by an ultra-thin hexagonal boron nitride (hBN) spacer, thereby reducing direct electronic hybridization while retaining strong Coulomb coupling [25]. Spacer-separated heterobilayers have emerged as a platform for long-lived dipolar interlayer excitons and collective bosonic phases [10,25], with recent work connecting these systems with a prospective exciton (quasi-) condensation as well as transport and localization phenomena [26–28]. While optical experiments have demonstrated a weakened overall moiré potential for the excitonic states in hBN-separated stacks [28–30], the implications of the spacer for any lattice reconstructions have not yet been studied. The atomic-scale structural role of the hBN layer is therefore directly relevant to efforts in exciton transport, the engineering of collective states, and the spontaneous formation of exciton coherence. As well as the effects of reconstruction, the effects of uniformity of the moiré potential are also important. In near-aligned TMD heterostacks, the structure remains insufficiently characterized to evaluate inhomogeneity of the moiré potentials across the length scales of optical experiments, i.e. micrometers. For instance, if an increased layer separation weakens the stacking-dependent interaction that drives atomic reconstruction, the spacer-separated region should remain closer to a rigid moiré superlattice. If a substantial relaxation persists, however, nominal twist angle and lattice mismatch will remain incomplete descriptors of the local stacking landscape, local rotation, and heterostrain. A direct structural comparison between direct-contact and spacer-separated regions is therefore essential.

This work combines high-angle annular dark-field scanning transmission electron microscopy (HAADF-STEM) with four-dimensional STEM (4D-STEM) to compare adjacent regions of a near-aligned, hBN-encapsulated $MoSe_2/WSe_2$ heterostack in which the TMD monolayers are either in direct contact or separated by monolayer or bilayer hBN spacers. Low-magnification and atomic-resolution HAADF-STEM reveal pronounced atomic reconstruction in the direct-contact region, whereas the spacer-separated region appears consistent with a rigid-moiré-like arrangement. Reciprocal-space analysis of 4D-STEM datasets further quantifies the corresponding distributions of local rotation, dilation, and heterostrain. The combined measurements show that an atomically thin hBN spacer strongly suppresses atomic reconstruction in near-aligned $MoSe_2/WSe_2$. The results motivate the use of monolayer spacers as a direct structural route toward a macroscopically uniform moiré landscape in van der Waals heterostructures and suggest that patterned spacer layers may present opportunities to create regions of spatially varying reconstruction, which is relevant to interlayer-exciton and correlated-boson physics.

## Results and Discussion

This study focuses on $MoSe_2/WSe_2$ heterobilayers that are in contact in some regions but separated by monolayer or bilayer hBN in other regions. These structures are encapsulated between top and bottom hBN flakes, each of thickness ≈ 15 layers (≈ 5 nm). The individual TMD monolayers are stacked in near-perfect alignment with a small twist angle of $\theta \approx 2°$, measured as described below. Figure 1(a) shows a schematic cross section of the sample. After fabrication, the heterostack is transferred from a $SiO_2/Si$ substrate to a TEM grid (see Experimental Section for details). The heterobilayer region extends over tens of micrometers and covers most of the TEM grid window, as shown in Figure 1(b). Direct-contact ($MoSe_2/WSe_2$) and spacer-separated ($MoSe_2$/hBN/$WSe_2$) regions with 1L- and 2L-hBN as a spacer are therefore both accessible within the same specimen.

We collect atomic-resolution HAADF-STEM images through the stack, using an accelerating voltage of 60 kV to mitigate beam damage. Figure 1 shows images obtained from the direct-contact and spacer-separated regions, as well as on the junction where they meet. The signal in these images is dominated by $Z$-contrast from the transition-metal sublattice, with $Z$ the atomic number ($Z_{Mo}$ = 42, $Z_{W}$ = 74). In the direct-contact region, even at low magnification, triangular domains are visible [Figure 1(c)]. The formation of such domain patterns is a hallmark of atomic reconstruction, in which the interlayer coupling gives rise to large areas of specific atomic alignments (atomic registries) between the constituent lattices. The reconstruction domains are anisotropically elongated across the field of view [Figure 1(c,d)], suggesting that a long-range strain field is present. Within the reconstructed region, the fast Fourier transform (FFT) exhibits a distorted hexagon of moiré peaks [white dashed box and inset in Figure 1(c)], indicating that the reconstruction periodicity differs along the three principal crystallographic directions. These inhomogeneities and the strain of the reconstructed lattice will be quantified and discussed in more detail below.

In contrast, the region separated by a single hBN spacer layer shows a uniform superlattice pattern at low magnification that is consistent with a rigid-moiré-like arrangement [Figure 1(e,f)]. Correspondingly, the FFT of the full image yields a symmetric hexagonal arrangement of superlattice peaks in reciprocal space. The moiré period measured, $a_m \approx (8.5 \pm 0.3)\, nm$, confirms a twist angle close to 2°. Measurements taken on positions with the 2L-hBN spacer yield results that are similar to the 1L-hBN case, within the resolution of the HAADF images shown in Figure S1. For both 1L and 2L, the lattice reconstructions seen in the direct-contact region are not visible. Due to the similarity of 1L and 2L behaviour, the analysis described below focuses on measurements from regions with a single monolayer hBN spacer.

In order to probe the reconstruction in more detail, we obtain higher magnification HAADF-STEM images that resolve the local stacking registries of the two TMD lattices (Figure 2). In the direct-contact region, all three high-symmetry registries, $\mathrm{H_M^M}$ (blue), $\mathrm{H_X^M}$ (yellow), and $\mathrm{H_X^X}$ (pink), can be identified [Figure 2(a)] and are shown at higher magnification in Figure 2(c)-(e). However, these domains clearly have unequal sizes, forming a kagome-like contrast pattern in the direct contact region. The TMD monolayers are therefore reconstructed. We expect the interface to favor whichever registry has the lowest-energy atomic configuration. In this case, the reconstructed lattice is dominated by enlarged domains of the $\mathrm{H_X^M}$ registry, which is the one that is energetically favored for antiparallel (H-type) stacking [15]. The hBN-separated region shows dramatically different images. Here, the three registries are visible only at localized sites rather than forming larger or unequal reconstructed domains. These sites are shown as the blue, yellow and pink circles in Figure 2(b) and repeat periodically with the moiré cell. The images suggest that there is little to no reconstruction of the separated TMD monolayers.

To support the above interpretation of the contrast, the `abTEM` Python package is used to perform multislice HAADF-STEM simulations using the experimental imaging parameters [31]. First, the appearance of the three registries is simulated. It is not necessary to simulate the entire moiré cell, including the rotation between the lattices. Instead, these simulations include a $MoSe_2/WSe_2$ heterobilayer in perfect antiparallel alignment, and the three high-symmetry atomic configurations are calculated by shifting one lattice with respect to the other. The atomic model and simulated HAADF-STEM images are shown for each registry below the experimental data in Figure 2(c)-(e). Although the distinction between $\mathrm{H_M^M}$ and $\mathrm{H_X^X}$ can be subtle in projected HAADF-STEM contrast, the observation of extended $\mathrm{H_X^M}$ domains strongly supports antiparallel (H-type) alignment. In a parallel (R-type) bilayer, the analogous reconstructed registry would be $\mathrm{R_M^M}$, which is energetically unfavorable and therefore not expected to form comparably large domains [15]. The observed atomic-resolution contrast thus favors antiparallel alignment in our $MoSe_2/WSe_2$ heterobilayer.

These simulations replicating the three registries are complemented by a further simulation on a twisted heterobilayer with antiparallel alignment (Figure S2). This simulation is expected to replicate the entire high-magnification HAADF-STEM image of the spacer-separated region shown in Figure 2(b). The goal of the simulation is to test whether the registry of the moiré superlattice varies continuously in HAADF-STEM contrast. Due to computational constraints, the simulated lattice is limited to $20 \times 20$ unit cells of each material (2400 total atoms), which implies that the full periodicity of the rigid moiré superlattice is within the simulated structure if the twist angle is 4°. As seen in Figure S2(b), the simulation results reveal a moiré superlattice with triangular periodicity that is qualitatively consistent with the HAADF-STEM contrast visible in the spacer-separated region of the sample.

The HAADF-STEM images shown so far successfully resolve the real-space domain morphology at the atomic scale. Nonetheless, the resulting image contrast is an indirect projection convoluted by out-of-plane lattice corrugation and inevitable amorphous surface residues [32]. To directly map the in-plane reconstruction landscape at a larger scale, we obtain convergent beam electron diffraction (CBED) data at an array of locations on the sample (i.e. 4D-STEM). Unlike HAADF-STEM, CBED collects coherently scattered electrons and captures the coherent phase interference of the diffracted electron waves [32,33]. By performing Bragg interferometry, the hexagonal lattice symmetry of a TMD crystal yields six Bragg discs in the detection plane. Figure 3(a) displays a simulated diffraction pattern averaged across a rigid $MoSe_2$/$WSe_2$ bilayer twisted by the experimental value of 2°. In the relevant case of a near-aligned bilayer with negligible lattice mismatch, the Bragg discs of the two monolayers overlap, thus creating six overlap regions. Under the weak phase object approximation, the intensity $I_j(\mathbf{r})$ of a $j$-th overlap region at a given real-space probe position $\mathbf{r}$ is modulated by the in-plane displacement $\mathbf{u}(\mathbf{r})$, as described by

$$I_j(\mathbf{r}) = A_j + B_j \cos\left(2\pi \mathbf{g}_j \cdot \mathbf{u}(\mathbf{r})\right) + C_j \sin\left(2\pi \mathbf{g}_j \cdot \mathbf{u}(\mathbf{r})\right),$$

with $\mathbf{g}_j$ the reciprocal lattice vectors [34]. Furthermore, the local relation between the reciprocal-lattice vectors $\mathbf{g}_j$ of the two monolayers is given by

$$\mathbf{g}_j^{WSe_2} = \mathbf{M}\mathbf{g}_j^{MoSe_2},$$

where the transformation matrix is written in polar form as $\mathbf{M} = \mathbf{RU}$, with $\mathbf{R}$ the rotational part and $\mathbf{U}$ the symmetric stretch part [35,36]. From $\mathbf{R}$, we obtain the local rotation after subtracting the mean twist, $\theta_R = \theta - \langle\theta\rangle$. Because $\mathbf{M}$ relates reciprocal-lattice vectors, the symmetric stretch obtained from the polar decomposition must be inverted to obtain the corresponding real-space stretch. The eigenvalues of this real-space stretch tensor yield the principal stretches $\lambda_1$ and $\lambda_2$, from which the principal strains are defined as $\epsilon_1 = \lambda_1 - 1$ and $\epsilon_2 = \lambda_2 - 1$. Then, the dilation is given by $e = \frac{\epsilon_1+\epsilon_2}{2}$, and the uniaxial heterostrain by $\epsilon_{\text{het}} = \frac{|\epsilon_1-\epsilon_2|}{2}$ [37].

Our measurements in reciprocal space intend to capture the lattice reconstruction defined by the displacement field encoded in the overlap intensity, as well as to quantify the structural parameters retrieved from the exact reciprocal-lattice vectors. Analysis with good resolution and over a large area is beneficial in quantifying the lattice distortion and evaluating its degree of uniformity and behavior at regions where the spacer layer thickness changes. We therefore obtained the 4D-STEM measurements using two different detectors. Data collected on an electron microscope pixel array detector (EMPAD, Thermo Fisher Scientific Inc.) was used for intensity-based Bragg interferometry over large areas, due to its efficient, fast read-out and high dynamic range (up to $10^6$:1). Data collected on a conventional CMOS camera (Ceta, Thermo Fisher Scientific Inc.), on the other hand, enables more accurate tracking of the layer-resolved Bragg-disc positions due to its larger pixel count, thereby enabling higher reciprocal-space resolution for more accurate measurements, albeit over smaller areas due to the slower speed. In this work, the EMPAD is employed for a large-area scan across the transition between the direct-contact and separated regions, while the Ceta data is used to extract and compare local rotation and strain fields in both regions of interest.

The main experimental parameters of the 4D-STEM scan with the EMPAD are an electron accelerating voltage of 60 kV, a beam current of ≈ 10-20 pA, a convergence semi-angle of 1.2 mrad, and an exposure time per pixel of 10 ms. A diffraction pattern averaged over all probe positions of the 4D-STEM scan is shown in Figure 3(b). As mentioned before, the two nearly aligned TMD lattices give rise to partially overlapping Bragg discs at six symmetry-equivalent positions in reciprocal space (yellow arrows). One representative region is enlarged in the red inset, where the individual $MoSe_2$ and $WSe_2$ discs are outlined by dashed circles. For each probe position $\mathbf{r}$, we quantify this signal by summing the intensity over the six overlap regions $\Omega_j$ in reciprocal space,

$$I_{overlap}(\mathbf{r}) = \sum_{j=1}^{6} \int_{\Omega_j} I(\mathbf{k}, \mathbf{r})\, d^2k$$

where $I(\mathbf{k}, \mathbf{r})$ is the diffraction intensity recorded at scattering vector $\mathbf{k}$ and probe position $\mathbf{r}$ and $j$ is the index of the Bragg disc [32]. The resulting overlap-intensity map is shown in Figure 3(c). Unlike the low-magnification HAADF-STEM images, the local in-plane interlayer displacement (representing the local stacking registry) is well encoded into the diffracted intensity. Consequently, the reconstruction pattern is well-defined by sharp boundaries between domains of distinct atomic registries. Within the mapped sample area, there is a clear boundary between the spacer-separated region on the left and the direct-contact region on the right (marked by the black arrow at the top). The direct-contact region displays varying atomic reconstruction and strongly elongated stripe-like domains, which become progressively more anisotropic toward the boundary and are consistent with the real-space morphology suggested by HAADF-STEM. By contrast, in the spacer-separated region, $I_{overlap}(\mathbf{r})$ exhibits a regular triangular modulation that closely matches the simulation of a rigid moiré superlattice for an H-type $MoSe_2/WSe_2$ bilayer twisted by 2°, as shown in the inset (see Experimental Section for simulation details). This agreement indicates that the spacer-separated region remains close to a rigid-moiré-like arrangement. Therefore, within this field of view, the reciprocal-space data from 4D-STEM provides evidence that the atomically thin hBN spacer prevents atomic reconstruction and preserves a nearly rigid moiré superlattice. Additionally, the elongated and highly anisotropic reconstruction pattern at the edge to the separated region suggests that all large-scale strain is accommodated by the lattice reconstructions, rather than influencing the moiré-like superlattice beyond the boundary. By directly influencing the domain formation, this effect is expected to have a direct impact on the optoelectronic properties of a heterostructure [23].

Beyond the overlap intensity, the diffraction patterns in a 4D-STEM scan contain quantitative information on the local lattice transformation, as mentioned above, and therefore provide a direct way to assess the impact of the hBN spacer on the relaxation of stress and strain and on the overall structural homogeneity [37–39]. We show data acquired over two 50 nm × 50 nm areas using the high-resolution Ceta camera in Figure S3, where the Bragg disc overlap intensities $I_{overlap}$ are summed to create maps of a direct-contact region and a region separated by monolayer hBN. The results are consistent with those already shown in Figure 3(c) and its inset, where the direct-contact region displays extended domains typical of atomic reconstruction, while the spacer-separated region looks qualitatively similar to the rigid-moiré simulation. However, with the higher resolution of the reciprocal-lattice vectors $\mathbf{g}_j$, one can also obtain from these measurements an analysis of the structural parameters, i.e. rotation, dilation and strain. Figure 4 shows maps of these deformation parameters accompanied by histograms representing the distribution of values over all pixels in the area. In the directly contacted region, dilation and strain amount to $e^{\mathrm{MoSe2/WSe2}} \approx 0.25\%$ and $\epsilon_{\mathrm{het}}^{\mathrm{MoSe2/WSe2}} \approx 0.78\%$. The maps and corresponding histograms of $\theta_R$, $e$, and $\epsilon_{\mathrm{het}}$ exhibit broad distributions with standard deviations of $\sigma(\theta_R) \approx 0.062°$, $\sigma(e) \approx 0.151\%$ and $\sigma(\epsilon_{\mathrm{het}}) \approx 0.091\%$, that indicate strong local lattice inhomogeneity. By contrast, in the hBN-separated region all three distributions are two to three times narrower, consistent with a more spatially uniform structure. The distribution widths from both scans are summarized in Table 1, alongside values of a rigid-moiré simulation of $MoSe_2/WSe_2$ twisted by 2°. The improved homogeneity in the spacer-separated region is accompanied by a substantial reduction in the mean uniaxial heterostrain, from $\epsilon_{\mathrm{het}}^{\mathrm{MoSe2/WSe2}} \approx 0.78\ \%$ in the directly contacted region to $\epsilon_{\mathrm{het}}^{\mathrm{MoSe2/hBN/WSe2}} \approx 0.13\ \%$ in the separated one. This mitigation of strain can be combined with the abrupt change of reconstruction along the edge of the 1L-hBN spacer [cf. Figure 3(c)], suggesting the opportunity to create directions and regions of spatially varying strain profiles within a single TMD heterostack. Such spatially engineered bilayers might allow the formation of 1D and 0D landscapes suitable for exciton transport and localization experiments [40].

The black curves overlaid on the lower-row histograms of Figure 4 show the results of an analogous analysis of the rigid-moiré simulation. Their close agreement with the measured widths indicates that the spacer-separated region approaches the rigid-limit distribution. At the same time, the mean dilation of $e \approx 0.37\%$ in the spacer-separated region remains larger than in the simulation, suggesting a residual uniform dilation that is not captured by the ideal rigid model.

While the present analysis in reciprocal space focuses on the 1L-hBN spacer, it is conceivable that a 2L-hBN spacer can further improve strain and dilation, even though the spatial variations of all structural parameters match the rigid-moiré simulation, and therefore already appear almost fully suppressed by the monolayer spacer. Nonetheless, the present comparison adds a spacer-dependent analysis to the established picture of TMD moiré reconstruction. Previous theoretical and experimental studies have shown that low-angle TMD bilayers can relax away from the rigid-moiré geometry through stacking-domain formation and dislocation networks [14–16]. More recent 4D-STEM work has further shown that this relaxation can include spatial variations in local rotation and dilation [18]. Our measurements extend this picture by showing that an atomically thin hBN spacer substantially suppresses these reconstruction signatures in near-aligned $MoSe_2/WSe_2$. The comparison with rigid-moiré simulations further indicates that the spacer-separated region approaches the lower-bound distribution widths expected from the nominal lattice mismatch and twist angle. At the same time, the residual mean dilation suggests that uniform relaxation or sample-specific strain may remain even when spatial variations are strongly reduced. Since reconstruction and strain can modify the local moiré potential experienced by interlayer excitons, this improved structural homogeneity is relevant for the description of the moiré potentials as well as interpreting optical and transport measurements in near-aligned TMD heterostructures [19,21–23,41].

## Conclusion

In conclusion, HAADF-STEM and 4D-STEM measurements on adjacent direct-contact and hBN-separated regions of the same near-aligned, H-type $MoSe_2/WSe_2$ heterostack show that even a single layer of hBN spacer fully suppresses atomic reconstruction. Relative to the directly contacted bilayer, the spacer-separated region exhibits two to three times narrower distributions of local rotation, dilation, and uniaxial heterostrain. Furthermore, the mean uniaxial heterostrain is reduced by a factor of six. A comparison to the rigid-moiré simulations shows that the spacer-separated region approaches the rigid-moiré limit while retaining a small residual dilation offset. At the same time, the boundary to the separated region relaxes long-range strain and introduces quasi-1D lattice reconstructions.

These findings identify hBN spacers as a structural means of reducing reconstruction-driven inhomogeneity, in addition to their role in tuning interlayer coupling. This insight is directly relevant for experiments seeking more spatially homogeneous moiré heterostructures and for the interpretation of transport, coherence, and collective excitonic phenomena in near-aligned TMD bilayers. Moreover, the effect of strain concentrating at the boundary of the separated region establishes a novel pathway to the manipulation of strain fields in TMDs by spacer-layer engineering. It is to be expected that a plethora of two-dimensional materials, whether exfoliated or grown to desired shapes, could be exploited as spacer layers for innovative device architectures [42,43].

An important next step will be to extend the present in-plane analysis to out-of-plane degrees of freedom. In particular, phase-sensitive approaches based on Bragg-disc or convergent-beam electron holography could provide access to local corrugation, interlayer separation, and relative lattice distortions, while recently introduced 4D-STEM methods for topographic reconstruction suggest a route toward combining planar strain mapping with three-dimensional structural information [33,44]. Alternatively, deep-subangstrom phase retrieval via electron ptychography has emerged as a powerful computational approach to map the three-dimensional atomic coordinates and out-of-plane corrugations of twisted TMD bilayers [45], which could offer even more detailed information on the lattice relaxation as a function of spacer layer number. More broadly, correlative structural, optical, and transport measurements on the same heterostructure should make it possible to directly relate local moiré homogeneity and strain effects to the behavior of long-lived interlayer excitons and related many-body states.

## Experimental Section

### Sample Fabrication

Suitable hBN spacer layers were identified via optical contrast and atomic force microscopy (cf. Figure S4). The heterostack was assembled via mechanical exfoliation and transferred onto a $SiO_2$/Si substrate with the aid of a thin polycarbonate film [46]. The final stack was then annealed in vacuum for two hours at 150°C. A cellulose acetate butyrate (CAB) film was spin-coated onto the sample and with the aid of a liquid lift-off, the heterostack was subsequently transferred onto a 50 nm thick $Si_3N_4$ membrane window on a Si TEM grid (Norcada Inc., Edmonton, Alberta, Canada). The latter features a $25 \times 25$ array of 500 nm diameter sized holes at 1 µm pitch. The TEM grid also features a 5 nm carbon layer for improved charge dissipation during electron microscopy, deposited before flake transfer [47].

### Transmission Electron Microscopy

STEM imaging was performed using a probe-corrected Thermo Fisher Titan Themis Z operated at an electron accelerating voltage of 60 kV and a beam current of ≈ 14.8 pA. Detectors include an electron microscope pixel array detector (EMPAD, Thermo Fisher Scientific Inc., Waltham, MA, USA) and a CMOS camera (Ceta, Thermo Fisher Scientific Inc., Waltham, MA, USA). For HAADF-STEM the EMPAD detector angle range is 50 mrad to 200 mrad. 4D-STEM is performed at a beam convergence angle of 1.2 mrad and is collected with both detectors, as explained in detail above.

### Simulations

HAADF-STEM image simulations as shown in Figure 2 were performed using `abTEM`, a Python package for transmission electron microscopy simulations based on the multislice method [31]. Monolayer $MoSe_2$ and $WSe_2$ structures were extracted from .cif files of "The Materials Project" [48]. Since the goal is to merely replicate the domains of single atomic registry, the two monolayers of the H-type heterobilayer are not twisted. Instead, using fractional lattice coordinates, the $MoSe_2$ layer was translated relative to $WSe_2$ to realize the desired high-symmetry registries $\mathrm{H}_\mathrm{M}^\mathrm{M}$, $\mathrm{H}_\mathrm{X}^\mathrm{M}$, and $\mathrm{H}_\mathrm{X}^\mathrm{X}$. These multislice HAADF-STEM simulations were then carried out at an electron energy of 60 kV, with a probe semi-convergence angle of 30 mrad, an annular detector range of 50 mrad to 200 mrad, a slice thickness of 4 Å, and a probe-position sampling of 0.05 Å.

As mentioned in the main text, the additional simulation of a rigid moiré superlattice was performed on a 4° twisted bilayer, showcasing the effect of continuously varying atomic alignment in HAADF-STEM contrast.

4D-STEM simulations were implemented with the `abTEM` package, as well. The key differences to the HAADF simulations are an experiment-matching probe semi-convergence angle of 1.2 mrad and a probe-position sampling of 0.5 nm to 1 nm. The resulting diffraction patterns were then analyzed analogously to the experimental 4D-STEM datasets.

4D-STEM Data Analysis

Quantitative maps of local twist, dilation, and heterostrain were obtained from the 4D-STEM datasets using established nanobeam electron diffraction strain-mapping procedures [37–39]. At each probe position, overlapping Bragg discs associated with the $MoSe_2$ and $WSe_2$ lattices were isolated, and their centres were refined by least-squares fitting. The refined disc positions were converted into reciprocal-lattice vectors relative to the central beam. A best-fit local linear transformation was then determined between corresponding $MoSe_2$ and $WSe_2$ reciprocal-lattice vectors, describing the relative rotation and distortion of one lattice with respect to the other. Polar decomposition of this transformation separated the rotational contribution from the symmetric lattice distortion. The rotational component yielded the local twist angle, while the symmetric distortion was converted into real-space dilation and heterostrain. The resulting quantities were evaluated at each probe position to generate spatial maps and distributions for the direct-contact and hBN-separated regions.

## Supporting Information

Supporting Information is available from the Wiley Online Library or from the author.

## Acknowledgements

We acknowledge U. Wurstbauer and A. Knorr for fruitful discussions, and financial support from the German Science Foundation (DFG) via grant HO 3324/16-1, and the clusters of excellence MCQST (EXS-2111) and e-conversion (EXS-2089), and the priority program 2244 (2DMP) via HO3324/13-2. F.M.R., K.C. and J.K. acknowledge support from the US National Science Foundation under Trailblazer Engineering Impact Award No. 2421694. This work was performed in part using MIT.nano's Characterization Facilities. K.W. and T.T. acknowledge support from the CREST (JPMJCR24A5), JST and World Premier International Research Center Initiative (WPI), MEXT, Japan.

## Data availability statement

The data that support the findings of this study are available from the corresponding author upon reasonable request.

## Bibliography

[1] N. P. Wilson, W. Yao, J. Shan, and X. Xu, Excitons and emergent quantum phenomena in stacked 2D semiconductors, Nature **599**, 383 (2021).

[2] M. Brotons-Gisbert, B. D. Gerardot, A. W. Holleitner, and U. Wurstbauer, Interlayer and Moiré excitons in atomically thin double layers: From individual quantum emitters to degenerate ensembles, MRS Bulletin **49**, 914 (2024).

[3] H. Yu, G.-B. Liu, J. Tang, X. Xu, and W. Yao, Moiré excitons: From programmable quantum emitter arrays to spin-orbit–coupled artificial lattices, Sci. Adv. **3**, e1701696 (2017).

[4] E. M. Alexeev et al., Resonantly hybridized excitons in moiré superlattices in van der Waals heterostructures, Nature **567**, 81 (2019).

[5] K. L. Seyler, P. Rivera, H. Yu, N. P. Wilson, E. L. Ray, D. G. Mandrus, J. Yan, W. Yao, and X. Xu, Signatures of moiré-trapped valley excitons in MoSe2/WSe2 heterobilayers, Nature **567**, 66 (2019).

[6] K. Tran et al., Evidence for moiré excitons in van der Waals heterostructures, Nature **567**, 71 (2019).

[7] L. A. Jauregui et al., Electrical control of interlayer exciton dynamics in atomically thin heterostructures, Science **366**, 870 (2019).

[8] P. Rivera et al., Observation of long-lived interlayer excitons in monolayer MoSe2–WSe2 heterostructures, Nat Commun **6**, 6242 (2015).

[9] B. Miller, A. Steinhoff, B. Pano, J. Klein, F. Jahnke, A. Holleitner, and U. Wurstbauer, Long-Lived Direct and Indirect Interlayer Excitons in van der Waals Heterostructures, Nano Lett. **17**, 5229 (2017).

[10] M. M. Fogler, L. V. Butov, and K. S. Novoselov, High-temperature superfluidity with indirect excitons in van der Waals heterostructures, Nat Commun **5**, 4555 (2014).

[11] L. Sigl, F. Sigger, F. Kronowetter, J. Kiemle, J. Klein, K. Watanabe, T. Taniguchi, J. J. Finley, U. Wurstbauer, and A. W. Holleitner, Signatures of a degenerate many-body state of interlayer excitons in a van der Waals heterostack, Phys. Rev. Research **2**, 042044 (2020).

[12] C. Zhang, C.-P. Chuu, X. Ren, M.-Y. Li, L.-J. Li, C. Jin, M.-Y. Chou, and C.-K. Shih, Interlayer couplings, Moiré patterns, and 2D electronic superlattices in $MoS_2$ /$WSe_2$ hetero-bilayers, Sci. Adv. **3**, e1601459 (2017).

[13] H. Yoo et al., Atomic and electronic reconstruction at the van der Waals interface in twisted bilayer graphene, Nat. Mater. **18**, 448 (2019).

[14] V. V. Enaldiev, V. Zólyomi, C. Yelgel, S. J. Magorrian, and V. I. Fal'ko, Stacking Domains and Dislocation Networks in Marginally Twisted Bilayers of Transition Metal Dichalcogenides, Phys. Rev. Lett. **124**, 206101 (2020).

[15] A. Weston et al., Atomic reconstruction in twisted bilayers of transition metal dichalcogenides, Nat. Nanotechnol. **15**, 592 (2020).

[16] M. R. Rosenberger, H.-J. Chuang, M. Phillips, V. P. Oleshko, K. M. McCreary, S. V. Sivaram, C. S. Hellberg, and B. T. Jonker, Twist Angle-Dependent Atomic Reconstruction and Moiré Patterns in Transition Metal Dichalcogenide Heterostructures, ACS Nano **14**, 4550 (2020).

[17] E. Tillotson et al., Scanning Electron Microscopy Imaging of Twist Domains in Transition Metal Dichalcogenide Heterostructures, ACS Nano **18**, 34023 (2024).

[18] M. Van Winkle et al., Rotational and dilational reconstruction in transition metal dichalcogenide moiré bilayers, Nat Commun **14**, 2989 (2023).

[19] M. Richter, Theory of interlayer exciton dynamics in two-dimensional transition metal dichalcogenide heterolayers under the influence of strain reconstruction and disorder, Phys. Rev. B **109**, 125308 (2024).

[20] N.-L. Bathen, T. Deilmann, A. Senkić, H. Lambers, R. Dana, K. Watanabe, T. Taniguchi, F. M. Ross, J. Klein, and U. Wurstbauer, Precise Twist Angle Determination in Twisted $WSe_2$ via Optical Moiré Phonons, Nano Lett. **26**, 3745 (2026).

[21] J. Figueiredo et al., Laterally extended states of interlayer excitons in reconstructed MoSe2/WSe2 heterostructures, Npj Quantum Mater. **10**, 96 (2025).
[22] J. Hagel, S. Brem, J. A. Pineiro, and E. Malic, Impact of atomic reconstruction on optical spectra of twisted TMD homobilayers, Phys. Rev. Mater. **8**, 034001 (2024).
[23] S. Zhao et al., Excitons in mesoscopically reconstructed moiré heterostructures, Nat. Nanotechnol. **18**, 572 (2023).
[24] L. H. Fowler-Gerace, Z. Zhou, E. A. Szwed, D. J. Choksy, and L. V. Butov, Transport and localization of indirect excitons in a van der Waals heterostructure, Nat. Photon. **18**, 823 (2024).
[25] Z. Wang, D. A. Rhodes, K. Watanabe, T. Taniguchi, J. C. Hone, J. Shan, and K. F. Mak, Evidence of high-temperature exciton condensation in two-dimensional atomic double layers, Nature **574**, 76 (2019).
[26] J. Cutshall et al., Imaging interlayer exciton superfluidity in a 2D semiconductor heterostructure, Sci. Adv. **11**, eadr1772 (2025).
[27] C. Lagoin and F. Dubin, Key role of the moir\'e potential for the quasicondensation of interlayer excitons in van der Waals heterostructures, Phys. Rev. B **103**, L041406 (2021).
[28] Z. Sun, A. Ciarrocchi, F. Tagarelli, J. F. Gonzalez Marin, K. Watanabe, T. Taniguchi, and A. Kis, Excitonic transport driven by repulsive dipolar interaction in a van der Waals heterostructure, Nat. Photon. **16**, 79 (2022).
[29] D. N. Shanks, F. Mahdikhanysarvejahany, T. G. Stanfill, M. R. Koehler, D. G. Mandrus, T. Taniguchi, K. Watanabe, B. J. LeRoy, and J. R. Schaibley, Interlayer Exciton Diode and Transistor, Nano Lett. **22**, 6599 (2022).
[30] Z. Li et al., Interlayer Exciton Transport in $MoSe_2$ /$WSe_2$ Heterostructures, ACS Nano **15**, 1539 (2021).
[31] J. Madsen and T. Susi, The abTEM code: transmission electron microscopy from first principles, Open Res Europe **1**, 24 (2021).
[32] C. Ophus, Four-Dimensional Scanning Transmission Electron Microscopy (4D-STEM): From Scanning Nanodiffraction to Ptychography and Beyond, Microsc Microanal **25**, 563 (2019).
[33] T. Latychevskaia, C. R. Woods, Y. B. Wang, M. Holwill, E. Prestat, S. J. Haigh, and K. S. Novoselov, Convergent beam electron holography for analysis of van der Waals heterostructures, Proc. Natl. Acad. Sci. U.S.A. **115**, 7473 (2018).
[34] M. Vulović, L. M. Voortman, L. J. Van Vliet, and B. Rieger, When to use the projection assumption and the weak-phase object approximation in phase contrast cryo-EM, Ultramicroscopy **136**, 61 (2014).
[35] T. C. Pekin, C. Gammer, J. Ciston, A. M. Minor, and C. Ophus, Optimizing disk registration algorithms for nanobeam electron diffraction strain mapping, Ultramicroscopy **176**, 170 (2017).
[36] B. H. Savitzky et al., py4DSTEM: A Software Package for Four-Dimensional Scanning Transmission Electron Microscopy Data Analysis, Microsc Microanal **27**, 712 (2021).
[37] N. P. Kazmierczak, M. Van Winkle, C. Ophus, K. C. Bustillo, S. Carr, H. G. Brown, J. Ciston, T. Taniguchi, K. Watanabe, and D. K. Bediako, Strain fields in twisted bilayer graphene, Nat. Mater. **20**, 956 (2021).
[38] V. B. Ozdol, C. Gammer, X. G. Jin, P. Ercius, C. Ophus, J. Ciston, and A. M. Minor, Strain mapping at nanometer resolution using advanced nano-beam electron diffraction, Applied Physics Letters **106**, 253107 (2015).
[39] Y. Han, K. Nguyen, M. Cao, P. Cueva, S. Xie, M. W. Tate, P. Purohit, S. M. Gruner, J. Park, and D. A. Muller, Strain Mapping of Two-Dimensional Heterostructures with Subpicometer Precision, Nano Lett. **18**, 3746 (2018).
[40] M. Heithoff et al., Valley-Hybridized Gate-Tunable 1D Exciton Confinement in $MoSe_2$, ACS Nano **18**, 30283 (2024).
[41] M. Kapfer et al., Programming twist angle and strain profiles in 2D materials, Science **381**, 677 (2023).

[42] C. Song et al., Growth and Interlayer Engineering of 2D Layered Semiconductors for Future Electronics, ACS Nano **14**, 16266 (2020).

[43] K. S. Kim et al., Non-epitaxial single-crystal 2D material growth by geometric confinement, Nature **614**, 88 (2023).

[44] A. Mireles, J. Park, S. H. Sung, C. Shi, B. Shin, J. Lou, C. Ophus, R. Hovden, K. Kang, and Y. Han, Strain mapping of three-dimensionally structured two-dimensional materials, Sci. Adv. **12**, eadz7908 (2026).

[45] K. X. Nguyen, Y. Jiang, C.-H. Lee, P. Kharel, Y. Zhang, A. M. Van Der Zande, and P. Y. Huang, Achieving sub-0.5-angstrom–resolution ptychography in an uncorrected electron microscope, Science **383**, 865 (2024).

[46] D. G. Purdie, N. M. Pugno, T. Taniguchi, K. Watanabe, A. C. Ferrari, and A. Lombardo, Cleaning interfaces in layered materials heterostructures, Nat Commun **9**, 5387 (2018).

[47] R. C. Williams and R. M. Glaeser, Ultrathin Carbon Support Films for Electron Microscopy, Science **175**, 1000 (1972).

[48] A. Jain et al., Commentary: The Materials Project: A materials genome approach to accelerating materials innovation, APL Materials **1**, 011002 (2013).

## Figures and Tables

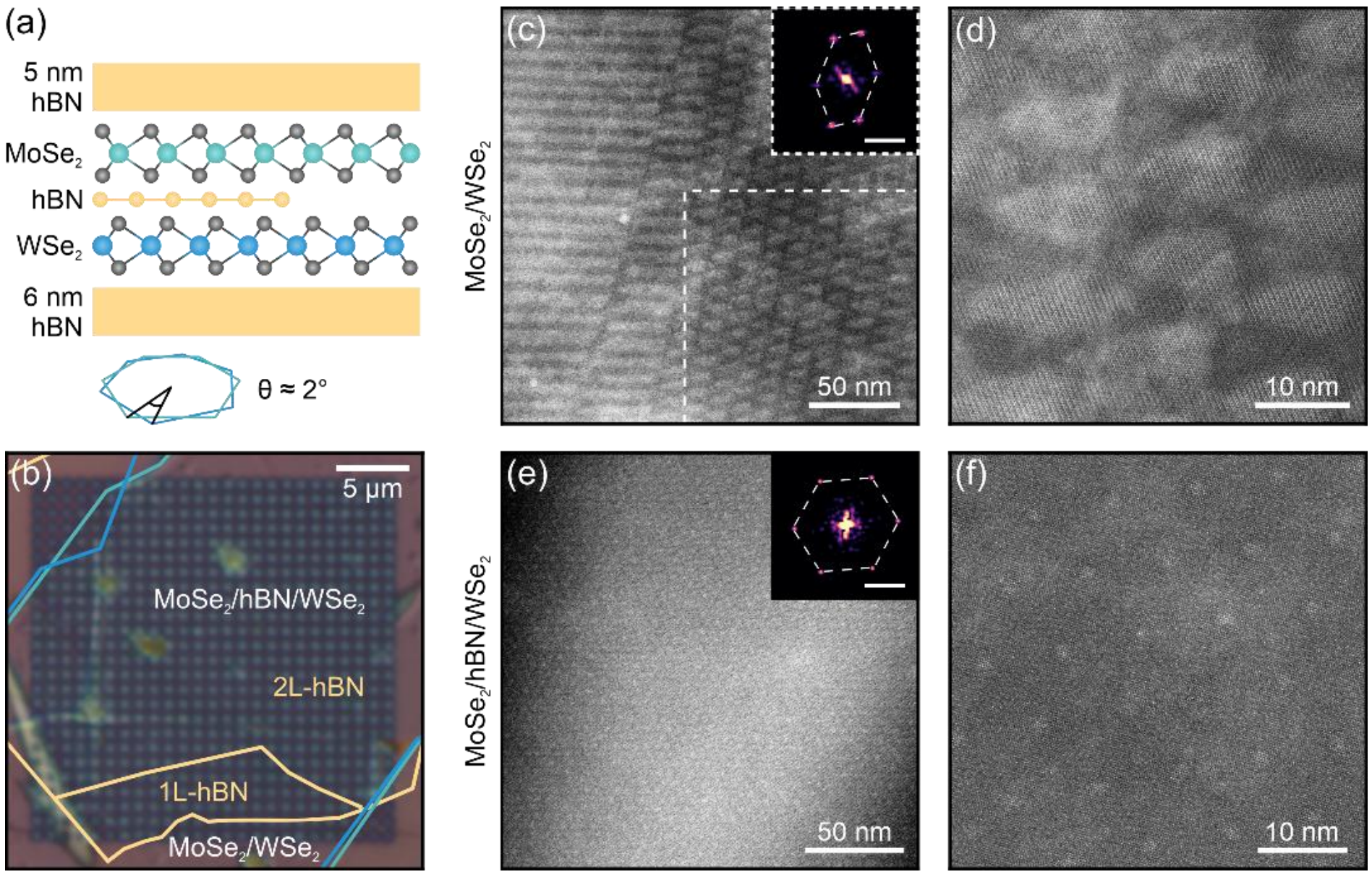

*Figure 1: Atomic ordering in directly contacted and spacer-separated $MoSe_2/WSe_2$ heterobilayers. (a) Schematic cross section of the heterostructure, consisting of an hBN-encapsulated $MoSe_2/WSe_2$ bilayer that is partially separated by monolayer or bilayer hBN. While the TMD bilayer is rotated by $\approx 2°$, the other layer rotations are random. (b) Optical micrograph of the sample. Scale bar, 5 µm. (c,d) Low- and medium-magnification HAADF-STEM images of the direct-contact region. The inset in (c) shows the FFT of the region outlined by the white dashed rectangle, yielding a distorted hexagonal arrangement of moiré peaks. Scale bar of inset, 0.1 $nm^{-1}$. (e,f) Same as in (c,d), but for the spacer-separated region. The corresponding FFT inset shows a nearly isotropic hexagonal arrangement of moiré peaks. Scale bar of inset, 0.1 $nm^{-1}$.*

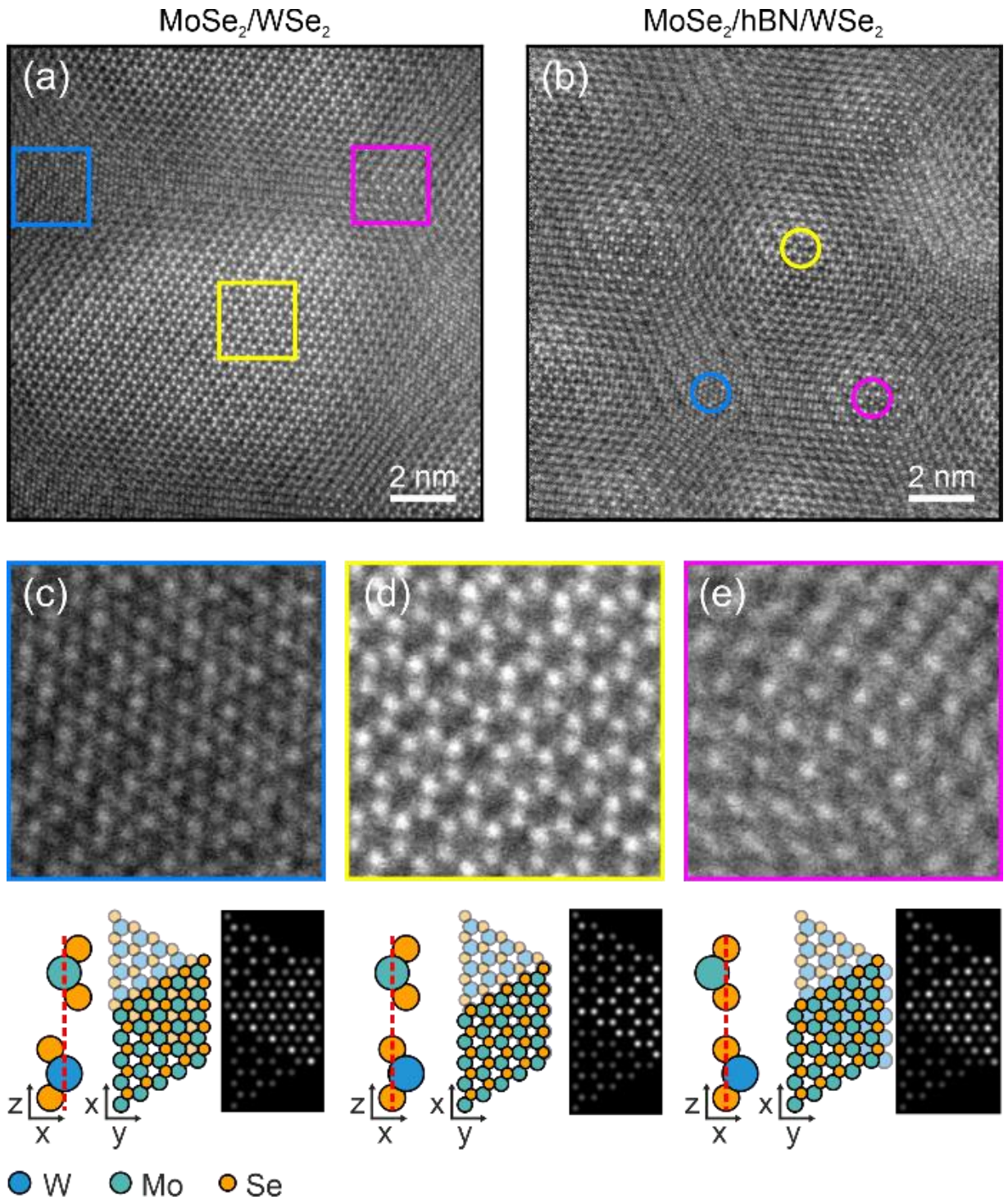

*Figure 2: Atomic resolution HAADF-STEM of $MoSe_2/WSe_2$. (a) High-magnification HAADF-STEM image in the direct-contact region. The atomic reconstruction pattern features domains consisting of three high-symmetry crystallographic alignments outlined by the colored boxes. The domains are separated by domain walls with continuously changing atomic registries. (b) Same as in (a), but in the spacer-separated region. The image reveals an undisplaced lattice with smoothly varying atomic registry, where the high-symmetry points, marked by colored circles, occupy a periodic triangular moiré superlattice. (c,d,e) Magnified cutouts as marked in (a). The three atomic registries $H_M^M$ (blue), $H_X^M$ (yellow), and $H_X^X$ (pink) are accompanied on the bottom by a schematic of the namesake atomic alignment between both layers and a simulated HAADF-STEM contrast of the displayed bilayer lattice. The red, dashed line highlights the vertical alignment of the atoms, respectively.*

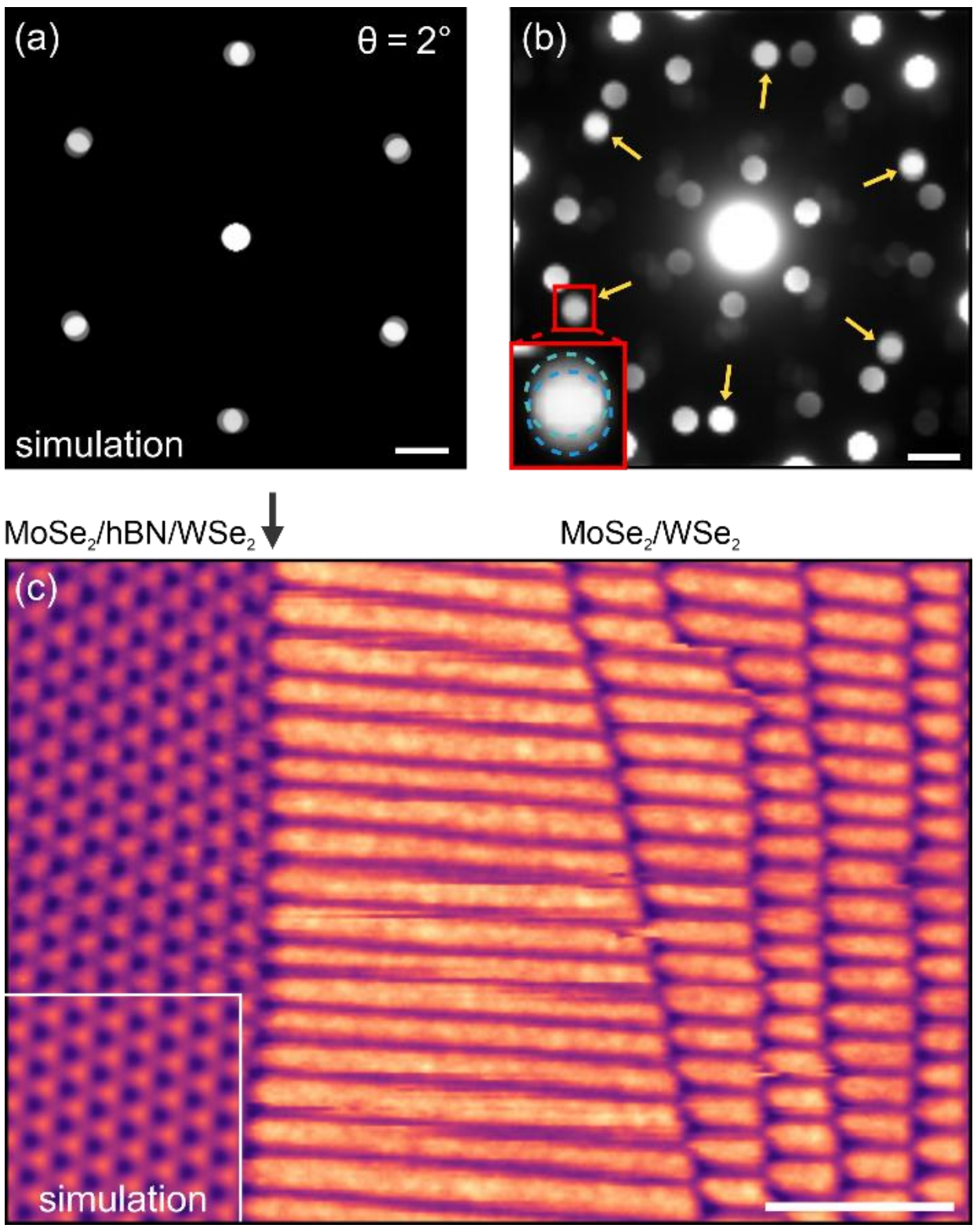

*Figure 3: 4D-STEM on spacer-separated and direct-contact $MoSe_2/WSe_2$. (a) Simulated diffraction pattern of an antiparallel rigid $MoSe_2/WSe_2$ bilayer twisted by ≈ 2°. Scale bar, 1 $nm^{-1}$. (b) Representative diffraction pattern. Yellow arrows mark the six symmetry-equivalent reciprocal-space positions where the $MoSe_2$ and $WSe_2$ Bragg discs overlap. The red inset shows an enlarged overlap region, with the two discs outlined by dashed circles. Scale bar, 1 $nm^{-1}$. (c) Real-space map obtained by summing the diffraction intensity in the six Bragg-disc overlap regions at each probe position. The black arrow marks the approximate boundary between the spacer-separated region on the left and the direct-contact region on the right. The spacer-separated region exhibits a regular triangular modulation consistent with a rigid moiré superlattice. The inset shows the result from the simulation with the experimental beam and detector parameters. In the direct-contact region, the overlap-intensity signal forms elongated stripe-like domains, consistent with anisotropic atomic reconstruction. Scale bar, 50 nm.*

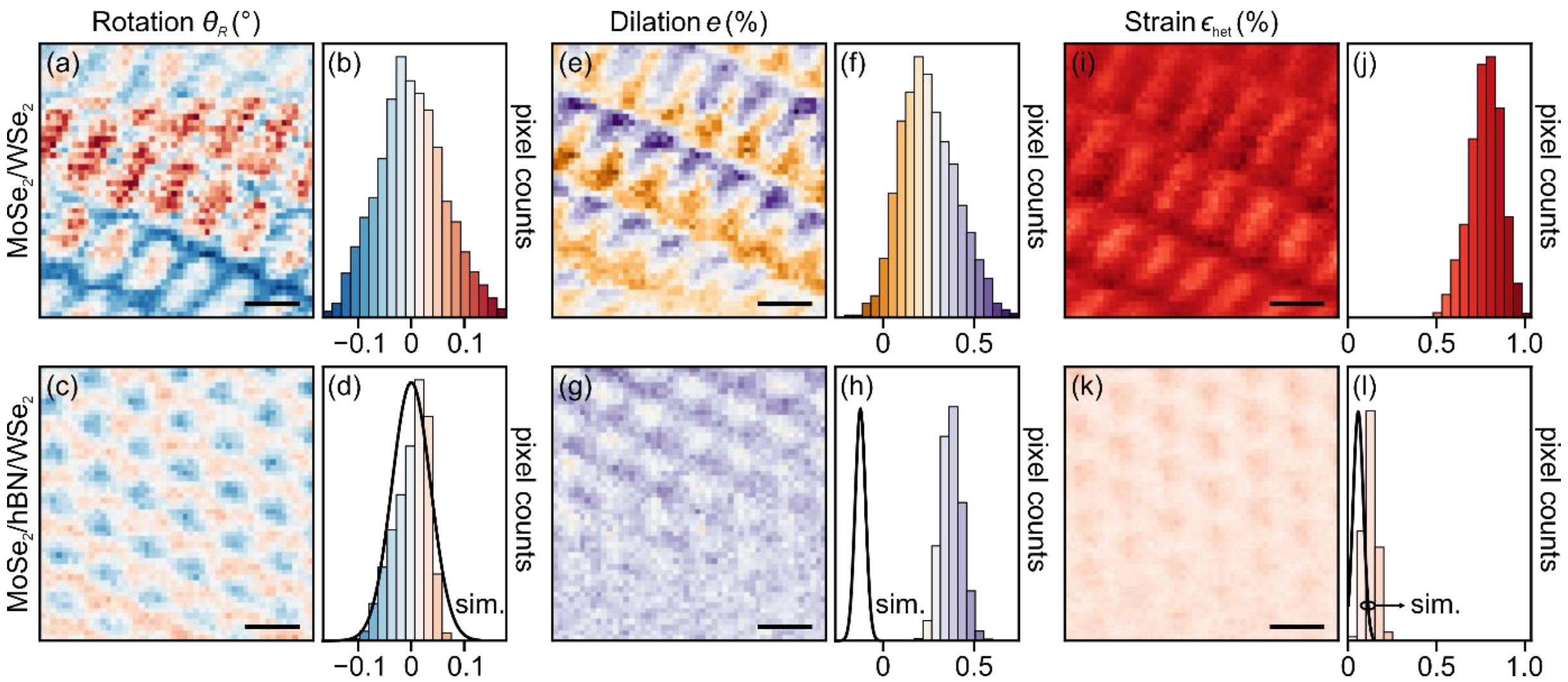

*Figure 4: Quantitative 4D-STEM analysis of the local lattice transformation in $MoSe_2/WSe_2$. (a,b) Real-space map and corresponding histogram of the local rotation after subtracting the mean twist, $\theta_R$, in the direct-contact region. (c,d) Same as in (a,b), but for the spacer-separated region. The black curve in (d) shows the corresponding distribution obtained from an analogous analysis of a simulated rigid moiré $MoSe_2/WSe_2$ bilayer. (e)--(h) Same as in (a)--(d), but for the dilation $e$. (i)--(l) Same as in (a)--(d), but for the uniaxial heterostrain $\epsilon_{het}$. For each quantity, the maps and histogram bins share the same color scale. Scale bars, 10 nm.*

*Table 1: Standard deviations of the distributions shown in Figure 4.*

| Region | $\sigma(\theta_R)$ | $\sigma(e)$ | $\sigma(\epsilon_{\text{het}})$ |
|---|---|---|---|
| Direct contact | 0.062° | 0.151% | 0.091% |
| hBN-separated | 0.032° | 0.059% | 0.035% |
| Rigid moiré simulation | 0.036° | 0.028% | 0.029% |

## Supporting Figures

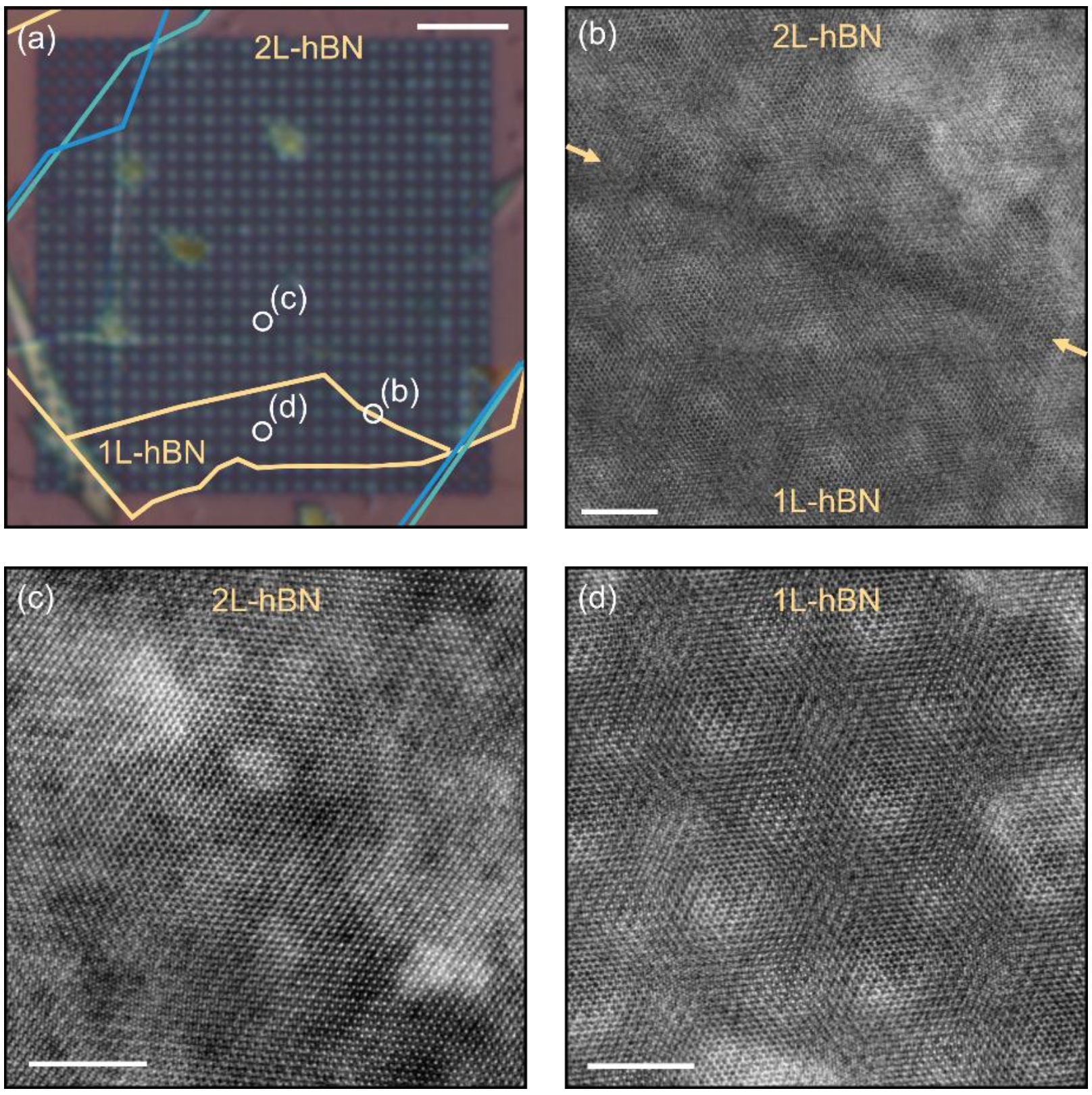

*Figure S1: (a) Optical micrograph of the TMD heterostack on the TEM grid. The white circles indicate the positions, where the subsequent HAADF-STEM images in (b)-(d) are collected from. Scale bar, 5 µm. (b) HAADF-STEM contrast at the step between 1L- and 2L-hBN separation. The yellow arrows mark the boundary. Scale bar, 5 µm. (c,d) HAADF-STEM contrast of area with a 2L-hBN and 1L-hBN spacer, respectively. Scale bars, 5 µm.*

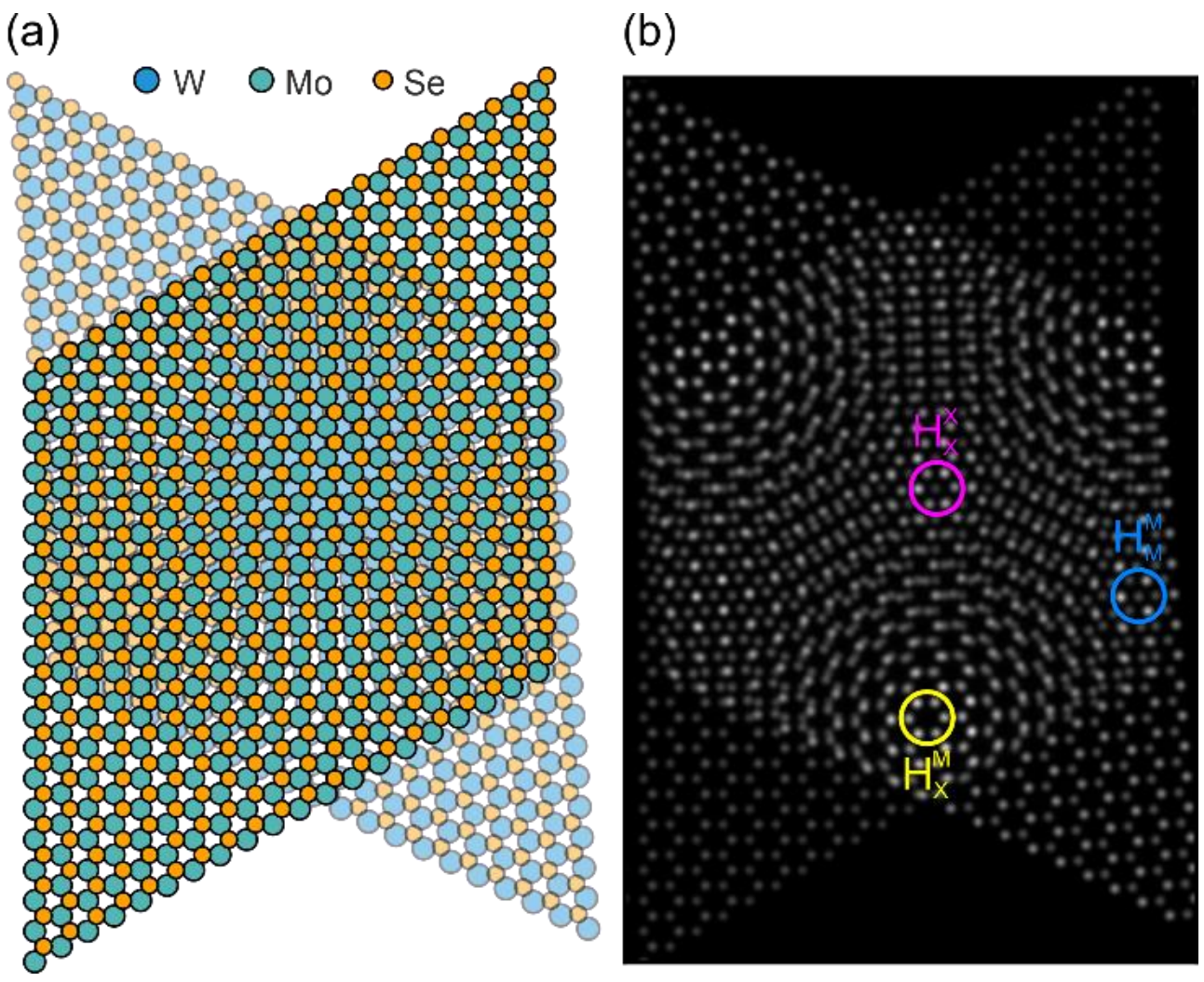

*Figure S2: (a) H-type $MoSe_2/WSe_2$ superlattice twisted by 4°. (b) HAADF-STEM contrast according to a multislice `abTEM` simulation with an annular detector range of 78 mrad to 200 mrad and a probe-position sampling of 0.1 Å. The high-symmetry points $H_M^M$ (blue), $H_X^M$ (yellow), and $H_X^X$ (pink) are highlighted with colored circles.*

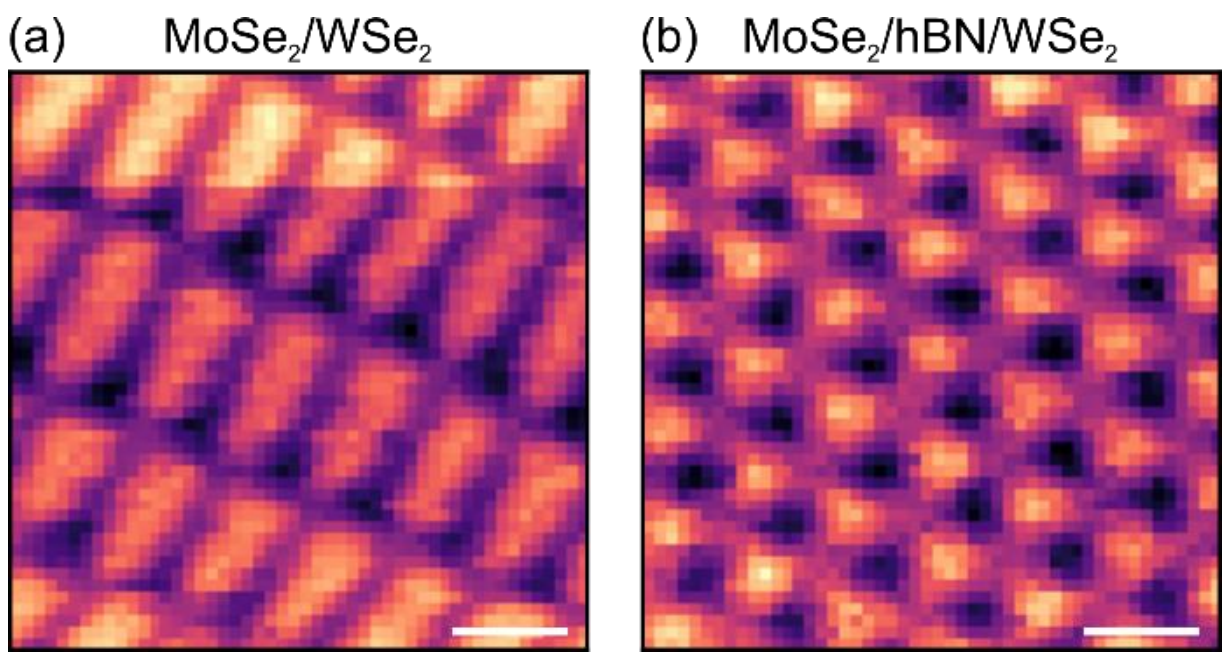

*Figure S3: (a) Real-space map of the Bragg disc overlap intensities $I_{overlap}$. The data is retrieved from the 4D-STEM scans analyzed in Figure 4. (b) Same as in (a) but for the spacer-separated region. Scale bars, 10 nm.*

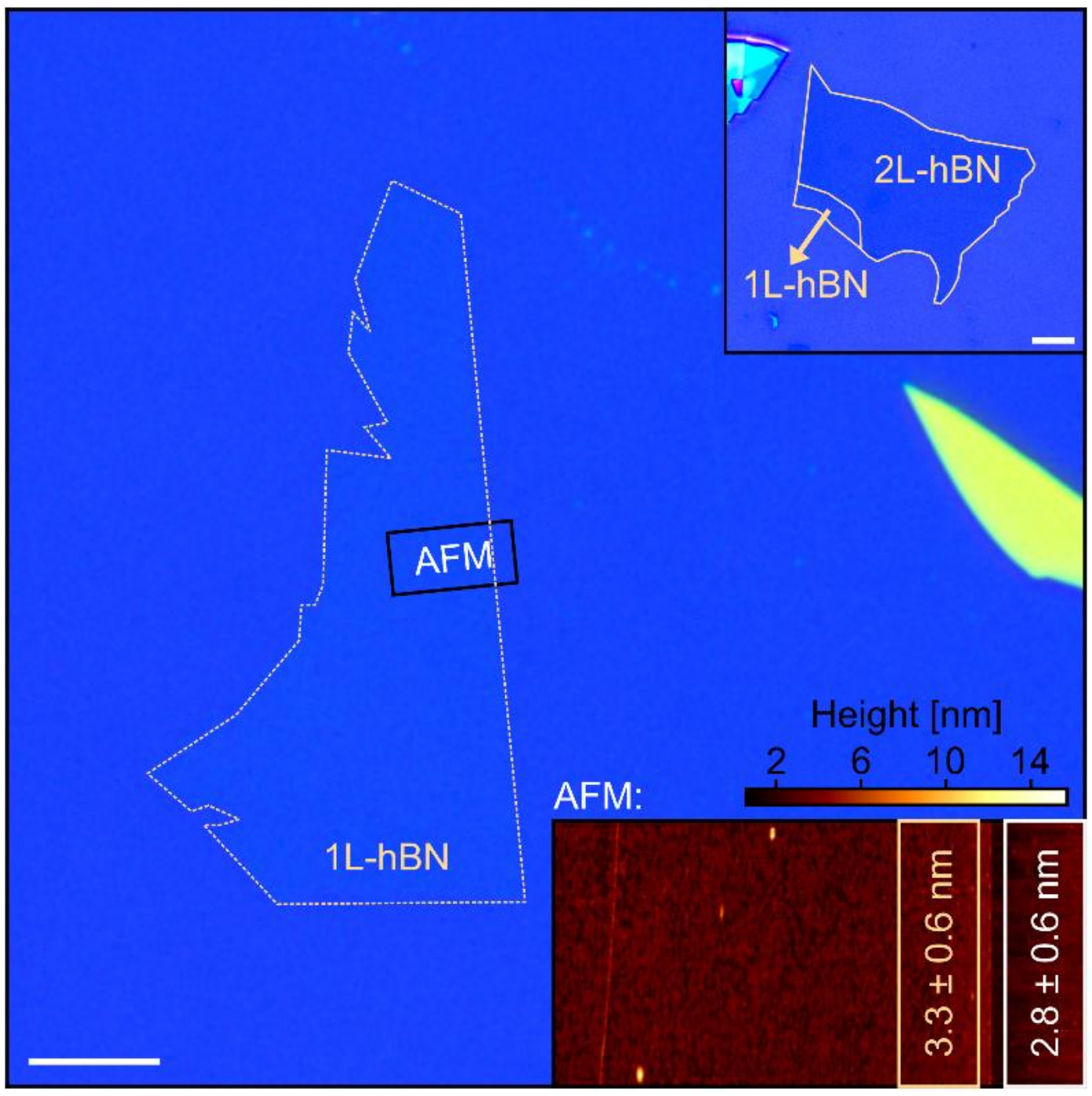

*Figure S4: Optical micrograph of a monolayer hBN flake on a silicon substrate with a 90 nm thick $SiO_2$ layer. The black box marks the 10 µm × 5 µm area of an AFM height scan. The AFM scan shown on the bottom right contains the average height of the scan areas left and right of the flake edge. The inset shows an optical micrograph of the hBN flake used in the analyzed sample. Scale bars, 10 µm.*

## Table of Contents Figure

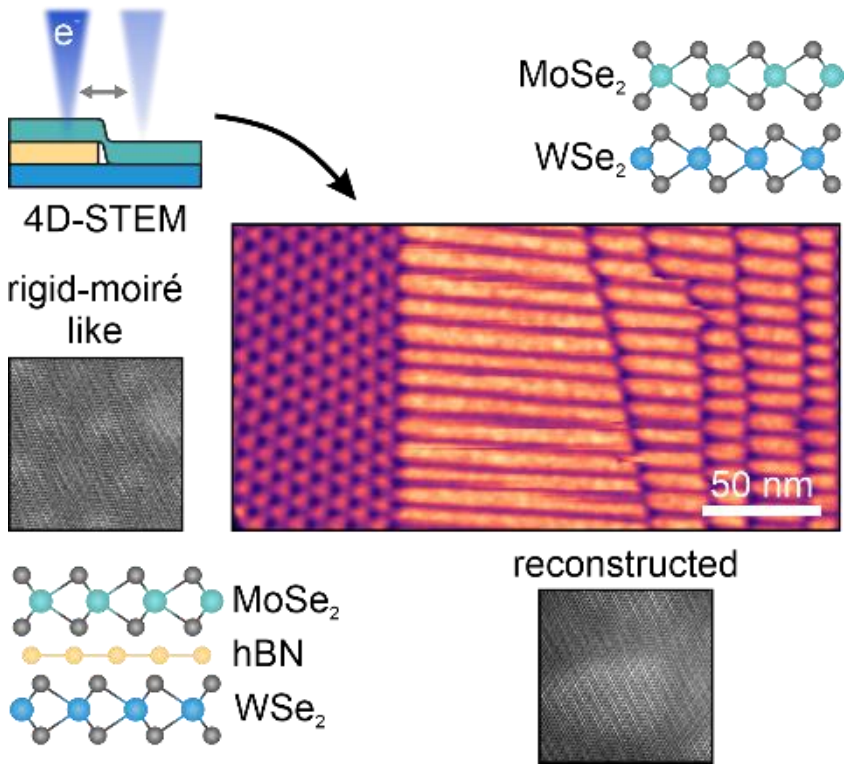

*ToC Figure: An ultrathin hexagonal boron nitride (hBN) spacer suppresses atomic reconstruction in near-aligned $MoSe_2/WSe_2$ heterobilayers. Both 4D-STEM and HAADF-STEM reveal that inserting a monolayer hBN spacer reduces local heterostrain by a factor of six and approaches ideal, rigid-moiré like behavior, providing a direct route toward structural homogeneity for macroscopical optoelectronic phenomena.*